\documentclass[pra,prl,showpacs,showkeys,floatfix,10pt]{revtex4-1}
\usepackage{graphicx}
\usepackage{dcolumn}
\usepackage{bm}

\usepackage{amsmath}
\usepackage{latexsym,amssymb,graphicx}
\usepackage[mathscr]{eucal}
\usepackage[bookmarks=true,
   colorlinks=true,
   linkcolor=blue,
   urlcolor=blue,
   citecolor=blue,
   bookmarks=true,
   hyperindex=true
]{hyperref}

\begin{document}


\title{Atomic Interferometric Gravitational-wave Space Observatory (AIGSO)}

\author{Dongfeng Gao\textsuperscript{1,2,}}
\altaffiliation{dfgao@wipm.ac.cn}
\author{Jin Wang\textsuperscript{1,2}}
\author{Mingsheng Zhan\textsuperscript{1,2,} }
\altaffiliation{mszhan@wipm.ac.cn}
\vskip 0.5cm
\affiliation{1 State Key Laboratory of Magnetic Resonance and Atomic and Molecular Physics, Wuhan Institute of Physics and Mathematics, Chinese Academy of Sciences - Wuhan National Laboratory for Optoelectronics, Wuhan 430071, China\\
2 Center for Cold Atom Physics, Chinese Academy of Sciences, Wuhan 430071, China }

\date{\today}

\begin{abstract}
We propose a space-borne gravitational-wave detection scheme, called atom interferometric gravitational-wave space observatory (AIGSO). It is motivated by the progress in the atomic matter-wave interferometry, which solely utilizes the standing light waves to split, deflect and recombine the atomic beam. Our scheme consists of three drag-free satellites orbiting the Earth. The phase shift of AIGSO is dominated by the Sagnac effect of gravitational-waves, which is proportional to the area enclosed by the atom interferometer, the frequency and amplitude of gravitational-waves. The scheme has a strain sensitivity $< 10^{-20}/\sqrt{{\rm Hz}}$ in the 100 mHz-10 Hz frequency range, which fills in the detection gap between space-based and ground-based laser interferometric detectors. Thus, our proposed AIGSO can be a good complementary detection scheme to the space-borne laser interferometric schemes, such as LISA. Considering the current status of relevant technology readiness, we expect our AIGSO to be a promising candidate for the future space-based gravitational-wave detection plan.

\end{abstract}

\maketitle

\section{Introduction}

An important prediction of Einstein's theory of general relativity is the existence of gravitational waves (GWs). The indirect evidence comes from the timing measurements of the binary pulsar, PSR 1913+16 \cite{taylor2010}. Since 1970s, a lot of ground-based GW detectors have been built to detect GWs directly \cite{ju2000}. After decades of efforts, the LIGO Scientific Collaboration and Virgo Collaboration announced the first two observations of GWs in 2016 \cite{ligo2016a, ligo2016b}, and three more observations in 2017 \cite{ligo2017a, ligo2017b, ligo2017c}, which opened a new era in the study of the universe. However, due to the seismic and Newtonian noise on the Earth, the ground-based laser interferometric detectors, such as LIGO, could only detect GWs with frequencies above 10 Hz. To search for GWs in the lower frequency band, several space-based laser interferometric detection schemes (such as LISA \cite{lisa2011}, TianQin \cite{tianqin2016}, DECIGO \cite{decigo2010}, BBO \cite{bbo2003}, and AMIGO \cite{ni2017}) have been proposed. However, to reach high sensitivity and wide operating frequency, these space-based laser interferometric GW detection proposals inevitably have to be large in size, and be expensive to build. Then, it is of great value to investigate alternative detection schemes that can have comparable sensitivity, be smaller in size, and be less expensive.

After more than twenty years of development, atom interferometers (AIs) have reached very high sensitivity \cite{cronin2009}. In fact, AIs have already been used in a large range of precision measurement experiments such as the measurements of the fine structure constant $\alpha$  \cite{clade2011} and the Newtonian gravitational constant $G$ \cite{tino2014}, which have already been adopted by the 2014 CODATA \cite{codata2014}. The AI was also used to give a $10^{-8}$-level test of the weak equivalence principle (WEP) in Ref. \cite{zhan2015}. Motivated by these impressive achievements, people began to work on the possibility of using AIs to detect GWs. Several terrestrial atom interferometric GW detection schemes were proposed in Refs. \cite{chiao2004,roura2006,tino2007,kasevich2008,gao2011}. Moreover, spaced-based atom interferometric GW detection schemes were put forward in Refs. \cite{chiao2004,kasevich2008,hogen2011,kasevich2016}. Compared to the space-borne laser interferometric GW detectors, the atoms are used as inertial test masses, instead of mirrors. Thus, to reach comparable sensitivity to LISA, these spaced-based atom interferometric GW detectors need to be of similar size too.

In this paper, we put forward a different proposal, called the atom interferometric gravitational-wave space observatory (AIGSO), which is an extension of our previous terrestrial GW detection scheme in Ref. \cite{gao2011}. Our AIGSO consists of three drag-free satellites, with a baseline length of 10 km, and a width of 1 m. With standing light waves, we can split, reflect and recombine a variety of atomic beams. What is more, the atoms always remain in the same internal state so that the atomic phase is not sensitive to external fields and fluctuations in laser beams. Our calculation shows that the phase shift of AIGSO is dominated by the Sagnac effect of GWs. In other words, our AIGSO is essentially an atomic Sagnac interferometric GW detector. Thus, it can be regarded as a matter-wave analogue of the laser Sagnac interferometric GW detector, which was considered as an attractive candidate for third-generation laser interferometric GW detectors in Ref. \cite{sun1996, chen2003}.

The remarking features of our AIGSO are the following. Compared to other spaced-based atom interferometric GW detectors in Refs. \cite{kasevich2008,hogen2011,kasevich2016}, our scheme is a genuine use of atomic matter-waves, not just used as inertial test masses instead. The size of our scheme is much smaller, and the atomic flux intensity can be much higher. Compared to the space-borne laser interferometric GW detectors, our scheme is much smaller in size, and is complementary in the GW frequency range. Our AIGSO is faster, where the integration time is reduced from one year to several days. Thus, in the same observation time, our scheme can detect GWs from more weaker sources.

The rest of the paper is organized as follows. In Sec. II, we give a description of our proposed scheme, and calculate the phase shift due to incoming GWs. The difference to other proposals can be clearly seen from the calculation. In Sec. III, we calculate the expected sensitivity of our detector. In Sec. IV, the potential detectable GW sources are discussed. Comments and conclusions are given in Sec. V.

\section{Our proposed AIGSO}

\subsection{Description of the AIGSO scheme}

\begin{figure}[htb]
\includegraphics[scale=0.7]{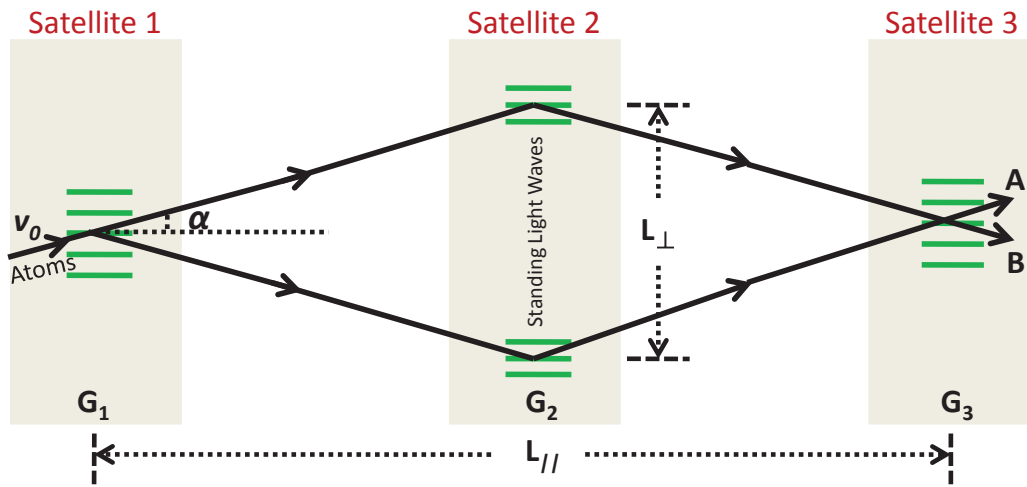}
\caption{Schematic diagram of our proposed AIGSO}
\label{fig1}
\end{figure}

The schematic diagram of our proposed AIGSO is depicted in FIG. \ref{fig1}, where the length and width of the interferometer are denoted by $L_{//}$ and $L_{\perp}$, respectively. Satellite 1 is used to host the atomic source and the first standing light wave. Satellite 2 hosts the middle two standing light waves. Satellite 3 houses the final standing light wave and the atom detection terminals A and B. All the standing light waves are connected by a common laser.

The atomic beam is first produced from a supersonic source, with initial velocity $v_0$. It should be well collimated so that the distribution of the transverse velocity of atoms can meet the requirement of GW detection. Then, the atomic beam is equally split into two beams of angle $2\alpha$, by the first standing light wave. After the propagation time, $\frac{T}{2} \equiv \frac{L_{//}}{2 v_0 cos\alpha}$ , the two atomic beams are reflected by the middle two standing light waves. After another propagation time $\frac{T}{2}$, the two beams are recombined, and go into the atom detection terminals A and B. Interference fringes can be formed in both terminals of the interferometer. Either of them can be used for GW detection because they are complementary to each other.

The size of our AIGSO is basically constrained by the environment outside the satellites since the atomic beams propagate between the satellites. The outside environment is full of background gases and solar photons. Collisions of atoms with them will destroy the coherence of atomic beams. According to the estimation in Ref. \cite{kasevich2008}, a space-borne atom interferometer, with interrogation time $T \leq 100$ s, sounds reasonable. For safety, let us take 10 s as the maximal interrogation time. The length of AIGSO is proposed to be $L_{//} = 10$ km. Then, the initial velocity $v_0$ of atomic beams should be $10^3$ m/s and higher. To minimize the outside environmental impact on the standing light waves, we want to operate them inside the satellites. Thus, the value of the width $L_{\perp}$ is also constrained by the size of satellites. Suppose the splitting angle $\alpha=10^{-4}$ rad, then the width $L_{\perp}$ is 1 m.

\subsection{The phase shift induced by GWs}

In this subsection, we will calculate the phase shift caused by the incoming GWs, which is very similar to the one for our terrestrial GW detection proposal in Ref. \cite{gao2011}. For completeness, we review the calculation as follows.

The interferometric area is supposed to be the $xy$ plane, and the GWs propagate in the $z$ direction. In the rigid frame, the metric for the GWs can be written as
\begin{equation}
ds^2=-dt^2+\delta_{ij}dx^idx^j+ (dz-dt) \dot{h}_{ij}\, x^idx^j ,
\end{equation}
where $z\equiv x^3$, $h^i_i=0$, $\partial^ih_{ij}=0$ and $i, j=1,2,3$. If the metric has $h_{11}=-h_{22}=h \, {\rm e}^{i (2\pi f t + \phi_0)}$, then it is called the $h_+$-polarization. According, the metric with $h_{12}=h_{21}=h \, {\rm e}^{i (2\pi f t + \phi_0)}$ is called the $h_\times$-polarization. $h$ and $f$ are the amplitude and frequency of the GW, respectively. $\phi_0$ is the initial phase of the GW at the time of entering into the interferometer.

The geodesic equations for particles confined in the $z=0$ plane are
\begin{equation}
\frac{d^2x^i}{dt^2}=\frac{1}{2}\ddot{h}^i_jx^j+\mathcal{O}(h^2_{ij},v^2h_{ij}).
\end{equation}
The Lagrangian for a non-relativistic atom is
\begin{equation}
L(x^i,\dot{x}^i)=\frac{m}{2}(\dot{x}^i\dot{x}_i-\dot{h}_{ij}\dot{x}^ix^j-2).
\label{lag}
\end{equation}
For a given trajectory, the phase at the final point can be calculated as
\begin{equation*}
\Phi_f=\Phi_i+\frac{1}{\hbar}S_{i\rightarrow f},
\end{equation*}
where $\Phi_i$ is the phase at the initial point, and $S_{i\rightarrow f}$ is the accumulated action along the trajectory, calculated from the Lagrangian (\ref{lag}).

Before the incoming of GWs, atoms move along the classical trajectories,
\begin{equation}
x_0(t)=v_0^x \, t \,, \,\,\,\,\,\,\,\,\,\,    y_0(t)= \left\{
\begin{array}{ll}
v_0^y \, t \,\,\,\,\,& {\rm for} \,\,\,\, 0<t<T/2 \\
v_0^y\, (T-t) \,\,\,\,\, & {\rm for}\,\,\,\, T/2<t<T \, ,
\end{array}\right.
\end{equation}
where $v_0^x=v_0 \cos \alpha\equiv V_{//}$, and $v_0^y= \pm v_0 \sin \alpha \equiv \pm v_{\perp}$. The $\pm$ signs are for the up arm and the down arm, respectively. In the present of GWs, both the atom's trajectory and velocity will be perturbed, which can be written as $\vec{x}=\vec{x}_0+\delta\vec{x} $ and $\vec{v}=\vec{v}_0+\vec{w}$.

Solving the geodesic equations for atoms, we can calculate the phase for each arm,
\begin{equation}
\frac{\hbar}{m}\Delta_{up,\, down}\Phi=\int^T_0 dt \left.\left[-\frac{(v_0^y)^2}{v_0^x}w^x+v_0^yw^y-\frac{1}{2}\dot{h}_{ij}v_0^ix_0^j\right]\right|_{up,\, down},
\end{equation}
where $\Delta_{up, \, down}\Phi$ means the phase change in the up and down arms, respectively.
Since the $h_+$-polarized GW affects the up and down arms in the same way, the net induced phase shift is zero. Our AIGSO scheme is only sensitive to the $h_\times$-polarization. By taking the difference of phase changes in the up and down arms, the phase shift is found to be
\begin{eqnarray}
\nonumber\frac{\hbar}{m}\Delta\Phi &=& -\frac{i \pi f h v_{\perp}(v_{//}-v_{\perp})T^2}{2} {\rm e}^{i \pi f T}{\rm e}^{i \phi_0}\\
            & & + \frac{ h T v_{\perp}({\rm e}^{i \pi f T}-1)}{v_{//}} \left[\frac{i (2 v_{//}^2-v_{\perp}^2)}{2 \pi f T}({\rm e}^{i \pi f T}-1)-\frac{v_{//}^2}{2}{\rm e}^{i \pi f T}+v_{\perp}^2\right]{\rm e}^{i \phi_0}.
\end{eqnarray}
Since $v_{\perp} \ll v_{//}$, the result can be further simplified into
\begin{eqnarray}
\nonumber\frac{\hbar}{m}\Delta\Phi &=& -\frac{i \pi f h v_{\perp}v_{//}T^2}{2} {\rm e}^{i \pi f T}{\rm e}^{i \phi_0}\\
            & & +  h T v_{\perp}v_{//}({\rm e}^{i \pi f T}-1) \left[(\frac{i}{\pi f T}-\frac{1}{2}){\rm e}^{i \pi f T}-\frac{i}{\pi f T}\right]{\rm e}^{i \phi_0}.
            \label{phase}
\end{eqnarray}
Note that $\Delta\Phi$ oscillates in time, because $\phi_0$ is oscillatory.

The feature of our result is that it contains two terms from two different contributions, which is quite similar to the result in Ref. \cite{tino2007}. Denote the first term in Eq. (\ref{phase}) by $\Delta\Phi_1$. We have $|\Delta\Phi_1|= \frac{m}{\hbar} 2\pi A f h$, where $A=\frac{1}{4}v_{//} v_{\perp} T^2$ is the area enclosed by the AIGSO. Then, it is clear that $\Delta\Phi_1$ is nothing but the GW-induced Sagnac phase shift, sensed by the atomic matte-waves. Denote the second term in Eq. (\ref{phase}) by $\Delta\Phi_2$. One has $|\Delta\Phi_2| \propto h L_{\perp}k_{dB}$, where $k_{dB}=m v_0/\hbar$ is the de Broglie wavenumber. Clearly, $\Delta\Phi_2$ is the well-known atomic matter-wave counterpart to the Michelson laser interferometric GW detector.

In the frequency range $f \ll 1/T$, the phase shift $\Delta\Phi$ is very small, and is dominated by $\Delta\Phi_2$ .  When $f \sim 1/T$, $\Delta\Phi_1$ and $\Delta\Phi_2$ are comparable. In the higher frequency range $f \gg 1/T$, the phase shift becomes bigger and is dominated by $\Delta\Phi_1$. In other words, our AIGSO has better sensitivity in the higher frequency band, where the laser interferometric GW detectors behave worse. Another feature of Eq. (\ref{phase}) is that $\Delta\Phi_1$ and $\Delta\Phi_2$ can not vanish simultaneously. Thus, the phase shift $\Delta\Phi$ is nonzero in the whole frequency range.

\section{Expected sensitivity of AIGSO}

Shot noise is the fundamental limit on the detection sensitivity of any interferometric GW detector. To compare the sensitivity of different GW detectors, people normally introduce the power spectral density for the shot-noise-limited sensitivity, $\tilde{h}_{sh} (f)$.  For our AIGSO, the $\tilde{h}_{sh} (f)$ can be easily found from Eq. (\ref{phase}),
\begin{eqnarray}
 \tilde{h}_{sh} (f)=\frac{2\hbar }{m|C(f)|\sqrt{\mathcal{R}}}
\end{eqnarray}
where
\begin{eqnarray}
C(f)&=&\frac{v_0^2 T \sin 2\alpha}{2} \left(\frac{\pi f T}{2} \sin(\pi f T) + \frac{\cos(\pi f T)- \cos(2\pi f T)}{2}-\frac{\sin(2\pi f T)- \sin(\pi f T)}{\pi f T}\right) \nonumber\\
&&+i \frac{v_0^2 T \sin 2\alpha}{2} \left(-\frac{\pi T f }{2} \cos(\pi f T) +\frac{1+\cos(2\pi f T)- 2\cos(\pi f T)}{\pi f T}-\frac{\sin(2\pi f T)- \sin(\pi f T)}{2}\right).
\end{eqnarray}
$\mathcal{R}$ is the flux intensity of the atomic beam.

This formula can be further simplified. Since the atomic beam is split by the standing light waves, the angle $2\alpha$ is written as
\begin{equation*}
\sin 2\alpha = \frac{p_{\perp}}{p_{//}} = \frac{N \hbar k_l}{m v_{//}} ,
\end{equation*}
where $p_{\perp}=N \hbar k_l$ is the momentum transferred from the standing light waves, $k_l=2\pi/\lambda_l$ is the wavenumber of the standing light waves, and $N$ is the number of photon momentum transfer to the atom. Eventually, the power spectral density for the sensitivity can be written into
\begin{eqnarray}
 \tilde{h}_{sh} (f)=\frac{\lambda_l }{\pi N L_{//} |\tilde{C}(f)|\sqrt{\mathcal{R}}}
\end{eqnarray}
where
\begin{eqnarray}
\tilde{C}(f)&=&\frac{1}{2} \left(\frac{\pi f T}{2} \sin(\pi f T) + \frac{\cos(\pi f T)- \cos(2\pi f T)}{2}-\frac{\sin(2\pi f T)- \sin(\pi f T)}{\pi f T}\right) \nonumber\\
&&+ \frac{i}{2} \left(-\frac{\pi T f }{2} \cos(\pi f T) +\frac{1+\cos(2\pi f T)- 2\cos(\pi f T)}{\pi f T}-\frac{\sin(2\pi f T)- \sin(\pi f T)}{2}\right).
\end{eqnarray}
We can see that $\tilde{h}_{sh} (f)$ is independent of the atomic mass, which is consistent with the WEP. It means that our scheme has the freedom in choosing the atomic species. Also, it is clear that the sensitivity of our AIGSO only depends on the parameters: $\lambda_l$, $N$, $L_{//}$, $T$, and $\mathcal{R}$.

To draw the sensitivity curve for our scheme, we have to specify these design parameters, which are supposed to be based on either current available technologies or future improvements. As discussed before, the size of AIGSO has been proposed to be $L_{//} = 10^4 \, {\rm m}$ and $L_{\perp} = 1\, {\rm m}$ to keep the coherence of the atomic matter-waves. The atomic beam is prepared from a supersonic source, with tunable initial velocity. For the case of $v_0=1000\, {\rm m\cdot s^{-1}}$, the interrogation time is then T=10 s.

To yield a good sensitivity in our scheme, the supersonic source of the atomic beam should have high intensity. Technologies of producing high intensity supersonic sources have been well developed \cite{scoles1992}. For example, Argon beams, with intensity $\mathcal{R}\sim {\rm 10^{19}\, atoms/s}$, was produced more than four decades ago \cite{leroy1969}. To be specific, in the following discussion, we will take Argon as our example. In our AIGSO, the intensity of Argon beam is proposed to be $\mathcal{R}\sim {\rm 10^{16} \, atoms/s}$. For a space-based detection scheme, this value of intensity should be realistic. For a one-year detection plan, the estimated consumption of atoms is about several moles. To manipulate the Ar atom, the wavelength of the standing light waves is $\lambda_l=810$ nm \cite{rasel1995}. To split the Ar beam by angle $\alpha=10^{-4}\, {\rm rad}$, we need a momentum transfer of $10 \hbar k_l$. Currently, momentum transfer of 102 $\hbar k_{l}$ has been reported \cite{kasevich2011a}. To split, reflect and recombine Ar beams of such an intensity, the estimated laser power in our scheme is less than one watt, which is lower than that of the laser interferometric GW detection plan in Ref. \cite{lisa2011}.

\begin{figure}[htb]
\includegraphics[scale=0.7]{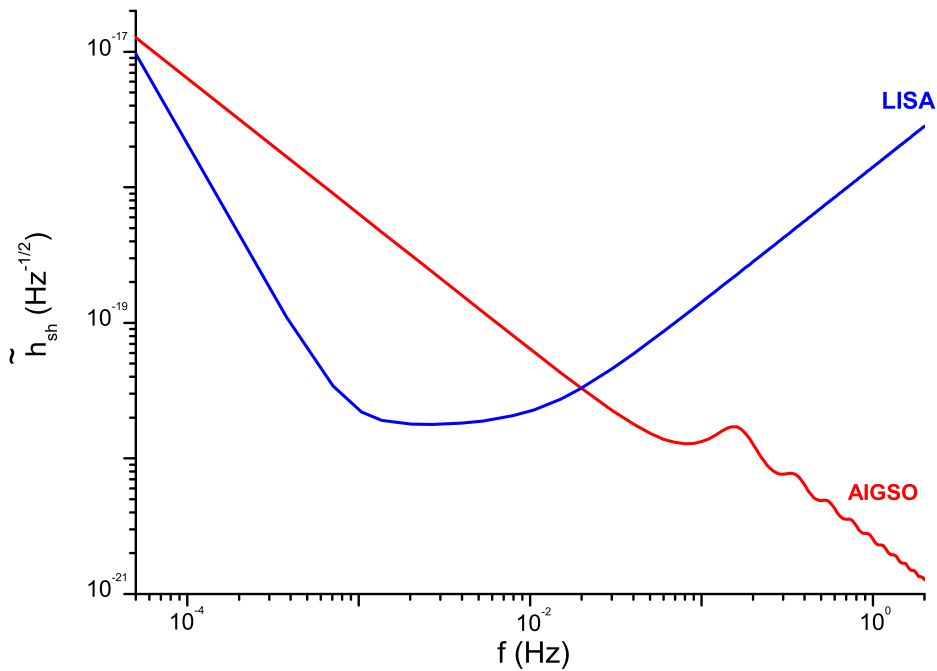}
\caption{Power spectral density for the shot-noise-limited sensitivity}
\label{fig2}
\end{figure}

With these proposed scheme parameters, the corresponding sensitivity curve for AIGSO is drawn in FIG.\ref{fig2}, where the curve for LISA is also shown for comparison. Our sensitivity curve is truncated at 1 Hz, which is the data taking rate $f_d$.  As discussed in Ref. \cite{kasevich2008}, $f_d$ is the frequency of running the atomic beam through the interferometer, which is no higher than 1 Hz ($\frac{10 s}{T}$). For our AIGSO, by tuning the interrogation time T, the data taking rate can be as high as 10 Hz. Compared to LISA, our AIGSO is worse at low frequencies. However, it works better and better at frequencies above 100 mHz, where LISA is not sensitive. Thus, our scheme could fill in the detection gap between space-based and ground-based laser interferometric GW detectors.

\section{GW sources for our AIGSO}

In the frequency band of AIGSO, the major sources of GWs are compact binaries, which consist of white dwarfs (WDs), neutron stars (NSs), or black holes (BHs). They emit strong GW radiations in the high frequency band at the late stage of their evolution.

The GW radiation from the evolution of compact binaries can be well calculated in the quadrupole approximation \cite{peters1964, kopparapu2007}. The GW strain amplitude $h(f)$, from a circular binary of mass $M_1$ and $M_2$ with orbital frequency $f_{orb}$ (then the GW frequency $f_{GW}=2 f_{orb}$), is calculated to be
\begin{equation}
h(f)= \frac{16 \pi^2 G}{r c^4} \frac{f_{orb}^2 M_1 M_2 a^2}{M_{tot}},
\end{equation}
where r is the distance to the source, $M_{tot}=M_1 +M_2$, and $a$ is the radius of the binary. With Kepler's third law, $4\pi^2 f_{orb}^2=G M_{tot}/a^3$, we finally get
\begin{equation}
h(f)= \frac{4 \pi^{2/3} G^{5/3}}{r c^4} \frac{M_1 M_2}{M_{tot}^2} M_{tot}^{5/3} f_{GW}^{2/3} .
\end{equation}

For the purpose of GW detection, we also have to consider the time dependence of both the strain amplitude and the frequency. According to Ref. \cite{peters1964}, the time rate of change in the radius is
\begin{equation}
\dot{a}= -\frac{64  G^3}{5 c^5} \frac{M_1 M_2 M_{tot}}{a^3} .
\end{equation}
Then, one can get
\begin{equation}
\dot{f}= \frac{96 \pi^{8/3} G^{5/3}}{5 c^5} \frac{M_1 M_2}{M_{tot}^2} M_{tot}^{5/3} f_{GW}^{11/3} .
\end{equation}
Here, one can introduce the inspiral timescale, which characterizes the remaining lifetime of a binary before merger
\begin{eqnarray}
 \tau_{chirp}&\equiv & \frac{3 f_{GW}}{8 \dot{f}_{GW}}= \frac{5 c^5}{256 \pi^{8/3} G^{5/3}} \frac{M_{tot}^2}{M_1 M_2} M_{tot}^{-5/3} f_{GW}^{-8/3} \nonumber \\
&=& 4.4\times 10^3 {\rm yr} \left(\frac{M_{tot}^2}{M_1 M_2}\right)\left(\frac{M_{tot}}{M_0}\right)^{-5/3} \left(\frac{f_{GW}}{10 \, {\rm mHz}}\right)^{-8/3} ,
\label{chirp}
\end{eqnarray}
where $M_0$ is the solar mass.

For a given GW source of strain amplitude $h(f)$, the detectability is determined by its corresponding strain density $\tilde{h}(f)$ in the detector. As discussed in Ref. \cite{hogen2011,hohensee2011}, $\tilde{h}(f)$ is given by
\begin{equation}
\tilde{h}(f)=h(f) \sqrt{\tau},
\end{equation}
where $\tau$ represents the shorter of the inspiral time $\tau_{chirp}$ and the observation time $\tau_{obs}$.

Assume that the observation time $\tau_{obs}$ of AIGSO is one year. From Eq. (\ref{chirp}), we can see that the inspiral time $\tau_{chirp}$ is strongly constrained by the masses of a binary and the frequency of GWs it radiates. For WD-WD binaries, due to their relatively large size, the mergers end at frequency below 1 Hz. So the inspiral time $\tau_{chirp}$ is greater than $\tau_{obs}$. For NS-NS binaries, $\tau_{chirp}$ varies from several thousand years at 10 mHZ to several hundred seconds at 10 Hz. After that, they enter the frequency domain of ground-based laser interferometric GW detectors, such as Advanced LIGO. For more massive binaries, such as intermediate massive BHs, $\tau_{chirp}$ is shorter than one year.

\begin{figure}[htb]
\includegraphics[scale=0.7]{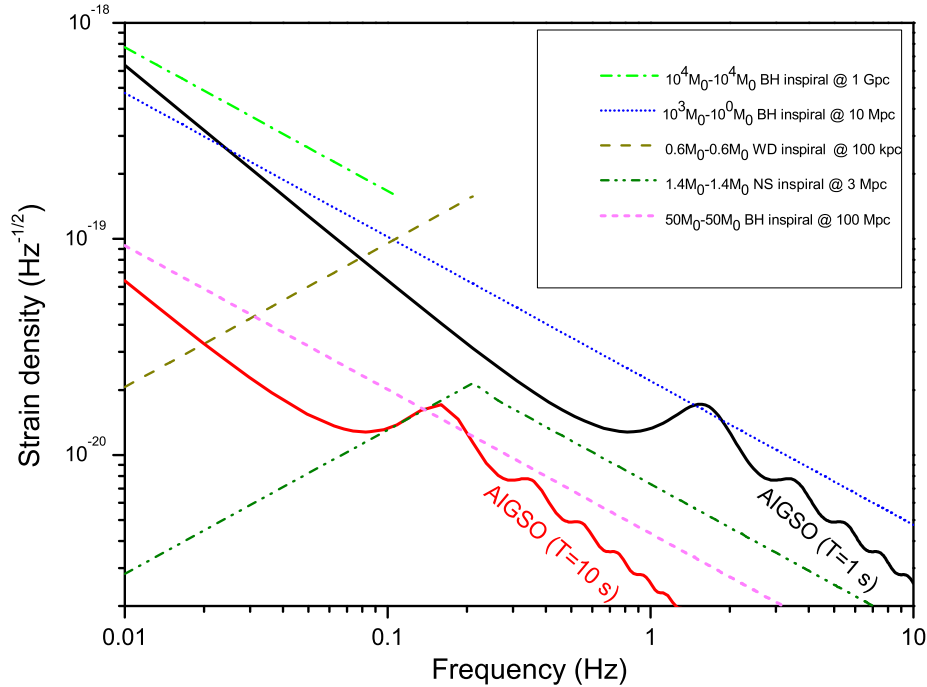}
\caption{Strain density for several potential GW sources. A signal-to-noise ratio of 5 is assumed. For comparison, the sensitivity curves for AIGSO with T=1 s and 10 s are also shown.}
\label{fig3}
\end{figure}

In FIG. \ref{fig3}, the strain density $\tilde{h}(f)$ curves for several detectable GW sources are drawn. Our AIGSO is sensitive to GWs from WD-WD binaries as far as 100 kpc. According to the analysis of \cite{kopparapu2007, hollberg2008}, there are plenty of such binaries around our galaxy. For NS-NS binaries, our AIGSO can detect GWs from them at a distance up to 3 Mpc. The bad thing is that NS-NS binaries are not so common. At the optimistic likely rate, one can hope to detect some \cite{abadie2010}. Lastly, the most important sources are BH-BH binaries. Our scheme can detect $10^4 M_0$-$10^4 M_0$ BH inspirals at 1 Gpc, and $10^3 M_0$-$10^0 M_0$ BH inspirals at 10 Mpc. Especially, our scheme is also sensitive to the $M_{tot} \leq 100 M_0$ BH inspirals at 100 Mpc, which are the main targeted sources for Advanced LIGO \cite{abbott2016c}. Since intermediate massive BHs are not well studied, the estimated likely rate is not so certain \cite{abadie2010}.

\section{Conclusions}

Through this preliminary study, we have described our atomic matter-wave Sagnac interferometric GW detection scheme, AIGSO, consisting of three drag-free satellites. Unlike other spaced-based atom interferometric GW detection proposals, our scheme is a genuine use of atomic matter-wave interferometry, instead of using just as inertial test masses. Compared to the space-borne laser interferometric GW detectors, our AIGSO is about 5 orders of magnitude smaller in size, which hopefully means the cut-down in technological requirements and in expense. Just like the laser Sagnac interferometer, which has been considered as an attractive candidate for the third-generation terrestrial GW detectors, our scheme is conceptually new, and can be a promising candidate for the future space-based GW detector. Of course, before becoming a real project, we need a thorough analysis of our AIGSO, which will be done in our future work.

Our AIGSO has a strain sensitivity $< 10^{-20}/\sqrt{{\rm Hz}}$ in the 100 mHz-10 Hz frequency band, which just covers the blank detection range between space-based and ground-based laser interferometric detectors. Potential GW sources for our proposed scheme include: the WD-WD binaries at 100 kpc, the NS-NS binaries within 3 Mpc, the $10^4 M_0$-$10^4 M_0$ BH inspirals up to 1 Gpc, the $10^3 M_0$-$10^0 M_0$ BH inspirals at 10 Mpc, and the $M_{tot} \leq 100 M_0$ BH inspirals within 100 Mpc. Especially, the last type of sources are also the main targeted sources for current terrestrial GW detectors. These sources spend much more time in our frequency band before stepping into Advanced LIGO's frequency band. Thus, our AIGSO can serve as a messenger to the ground-based GW detectors, and finally join together to provide a full scan of GWs.

\begin{center}
\large{{\bf Acknowledgements}}
\end{center}

We would like to thank Wei-Tou Ni for many useful discussions. This work was supported by the National Key Research Program of China under Grant No. 2016YFA0302002, the National Science Foundation of China under Grants No. 11227803 and No. 91536221, and the Strategic Priority Research Program of the Chinese Academy of Sciences under Grant No. XDB21010100.


\begin{thebibliography}{99}

\bibitem{taylor2010}
J. M. Weisberg, D. J. Nice, and J. H. Taylor, Timing Measurements of the Relativistic Binary Pulsar PSR B1913+16, Astrophys. J. {\bf 722}, 1030 (2010).
\bibitem{ju2000}
L. Ju, D.G. Blair, and C. Zhao, Detection of gravitational waves, Rep. Prog. Phys. {\bf 63}, 1317 (2000), and references therein.
\bibitem{ligo2016a}
B. P. Abbott \textit{et al.} (LIGO Scientific Collaboration and Virgo Collaboration), Observation of gravitational waves from a binary black hole merger, Phys. Rev. Lett. {\bf 116}, 061102 (2016).
\bibitem{ligo2016b}
B. P. Abbott \textit{et al.} (LIGO Scientific Collaboration and Virgo Collaboration), GW151226: Observation of gravitational waves from a 22-solar-mass binary black hole coalescence, Phys. Rev. Lett. {\bf 116}, 241103 (2016).
\bibitem{ligo2017a}
B. P. Abbott \textit{et al.} (LIGO Scientific Collaboration and Virgo Collaboration), GW170104: Observation of a 50-Solar-Mass Binary Black Hole Coalescence at Redshift 0.2, Phys. Rev. Lett. {\bf 118}, 221101 (2017).
\bibitem{ligo2017b}
B. P. Abbott \textit{et al.} (LIGO Scientific Collaboration and Virgo Collaboration), GW170814: A Three-Detector Observation of Gravitational Waves from a Binary Black Hole Coalescence, Phys. Rev. Lett. {\bf 119}, 141101 (2017).
\bibitem{ligo2017c}
B. P. Abbott \textit{et al.} (LIGO Scientific Collaboration and Virgo Collaboration), GW170817: Observation of Gravitational Waves from a Binary Neutron Star Inspiral, Phys. Rev. Lett. {\bf 119}, 161101 (2017).
\bibitem{lisa2011}
K. Danzmann \textit{et al.}, LISA: Unveiling a hidden universe, Assessment Study Report ESA/SRE {\bf 3}, 2 (2011).
\bibitem{tianqin2016}
J. Luo \textit{et al.}, TianQin: a space-borne gravitational wave detector, Class. Quantum Grav. {\bf 33}, 035010 (2016).
\bibitem{decigo2010}
M. Ando \textit{et al.}, DECIGO and DECIGO pathfinder, Class. Quantum Grav. {\bf 27}, 084010 (2010).
\bibitem{bbo2003}
E. s. Phinney \textit{et al.}, NASA Mission Concept Study, 2003.
\bibitem{ni2017}
W.-T. Ni, Gravitational Wave (GW) Classification, Space GW Detection Sensitivities and AMIGO (Astrodynamical Middle-frequency Interferometric GW Observatory), arXiv: 1709.05659.

\bibitem{cronin2009}
A. D. Cronin, J. Schmiedmayer and D.E. Pritchard, Optics and interferometry with atoms and molecules, Rev. Mod. Phys. {\bf 81}, 1051 (2009), and references therein.
\bibitem{clade2011}
R. Bouchendira, P. Clad${\rm\acute{e}}$, S. Guellati-Kh${\rm\acute{e}}$lifa, F. Nez and F. Biraben, New determination of the fine structure constant and test of the quantum electrodynamics, Phys. Rev. Lett. {\bf 106}, 080801 (2011).
\bibitem{tino2014}
G. Rosi, F. Sorrentino, L. Cacciapuoti, M. Prevedelli and G. M. Tino, Precision measurement of the Newtonian gravitational constant using cold atoms, Nature (London) {\bf 510}, 518 (2014).
\bibitem{codata2014}
P. J. Mohr, D. B. Newell and B. N. Taylor, CODATA recommended values of the fundamental physical constants: 2014, Rev. Mod. Phys. {\bf 88}, 035009 (2016).
\bibitem{zhan2015}
L. Zhou, S. Long, B. Tang, X. Chen, F. Gao, W. Peng, W. Duan, J. Zhong, Z. Xiong, J. Wang, Y. Zhang and M. Zhan, Test of equivalence principle at $10^{-8}$ level by a dual-species double-diffraction Raman atom interferometer, Phys. Rev. Lett. {\bf 115}, 013004 (2015).


\bibitem{chiao2004}
R. Y. Chiao and A. D. Speliotopoulos, Towards MIGO, the matter-wave interferometric gravitational-wave observatory, and the intersection of quantum mechanics with general relativity, J. Mod. Opt. {\bf 51}, 861 (2004).
\bibitem{roura2006}
A. Roura, D. R. Brill, B. L. Hu, C. W. Misner and W. D. Phillips, Gravitational wave detectors based on matter wave interferometers (MIGO) are no better than laser interferometers (LIGO), Phys. Rev. D {\bf 73}, 084018 (2006).
\bibitem{tino2007}
G. M. Tino and F. Vetrano, Is it possible to detect gravitational waves with atom interferometers?, Class. Quantum Grav. {\bf 24}, 2167 (2007).
\bibitem{kasevich2008}
S. Dimopoulos, P. W. Graham, J. M. Hogan, M. A. Kasevich and S. Rajendram, An Atomic Gravitational Wave Interferometric Sensor (AGIS), Phys. Rev. D {\bf 78}, 122002 (2008).
\bibitem{gao2011}
D. Gao, P. Ju, B. Zhang and M. Zhan, Gravitational-wave Detection With Matter-wave Interferometers Based On Standing Light Waves, Gen. Rel. Grav. {\bf 43}, 2027 (2011).
\bibitem{hogen2011}
J. M. Hogan \textit{et al.}, An atomic gravitational wave interferometric sensor in low earth orbit (AGIS-LEO), Gen. Rel. Grav. {\bf 43}, 1953 (2011).
\bibitem{kasevich2016}
J. M. Hogan and M. A. Kasevich, Atom-interferometric gravitational-wave detection using heterodyne laser links, Phys. Rev. A {\bf 94}, 033632 (2016).
\bibitem{sun1996}
K. Sun, M. M. Fejer, E. Gustafson, and R. L. Byer, Sagnac Interferometer for Gravitational-Wave Detection, Phys. Rev. Lett. {\bf 76}, 3053 (1996).
\bibitem{chen2003}
Y. Chen, Sagnac interferometer as a speed-meter-type, quantum-nondemolition gravitational-wave detector, Phys. Rev. D {\bf 67}, 122004 (2003).

\bibitem{scoles1992}
G. Scoles, editor, {\it Atomic and Molecular Beam Methods}, Vols. 1, 2, Oxford University Press: New York, 1992.
\bibitem{leroy1969}
R. L. Leroy, T. R. Govers, and J. M. Deckers, Supersonic molecular beam intensities, Can. J. Chem. {\bf 48}(6), 927 (1970).
\bibitem{rasel1995}
E. M. Rasel, M. K. Oberthaler, H. Batelaan, J. Schmiedmayer, and A. Zeilinger, Atom wave interferometry with diffraction gratings of light, Phys. Rev. Lett. {\bf 75}, 2633 (1995).
\bibitem{kasevich2011a}
S. W. Chiow, T. Kovachy, H. C. Chien, and M. A. Kasevich, 102 $\hbar k$ Large Area Atom Interferometers, Phys. Rev. Lett. {\bf 107}, 130403 (2011).

\bibitem{peters1964}
P. C. Peters, Gravitational Radiation and the Motion of Two Point Masses, Phys. Rev. {\bf 136}, B1224 (1964).
\bibitem{kopparapu2007}
R. K. Kopparapu and J. E. Tohline, Population boundaries for galactic white dwarf binaries in LISA's amplitude-frequency domain, Astrophys. J. {\bf 655}, 1025 (2007).

\bibitem{hohensee2011}
M. Hohensee, S.-Y. Lan, R. Houtz, C. Chan, B. Estey, G. Kim, P.-C. Kuan, and H. M${\rm \ddot{u}}$ller, Sources and technology for an atomic gravitational wave interferometric sensor, Gen. Rel. Grav. {\bf 43}, 1905 (2011).

\bibitem{hollberg2008}
J. B. Holberg, E. M. Sion, T. Oswalt, G. P. McCook, S. Foran, and J. P. Subasavage, A New Look at the Local White Dwarf Population, Astronom. J {\bf 135}, 1225 (2008).
\bibitem{abadie2010}
J Abadie \textit{et al.}, Predictions for the rates of compact binary coalescences observable by ground-based gravitational-wave detectors, Class. Quantum Grav. {\bf 27}, 173001 (2010).
\bibitem{abbott2016c}
B. P. Abbott \textit{et al.}, Binary Black Hole Mergers in the First Advanced LIGO Observing Run, Phys. Rev. X {\bf 6}, 041015 (2016)


\end{thebibliography}
\end{document}